\documentclass[letterpaper]{article} % DO NOT CHANGE THIS
\usepackage{aaai2026}  % DO NOT CHANGE THIS
\usepackage{times}  % DO NOT CHANGE THIS
\usepackage{helvet}  % DO NOT CHANGE THIS
\usepackage{courier}  % DO NOT CHANGE THIS
\usepackage[hyphens]{url}  % DO NOT CHANGE THIS
\usepackage{graphicx} % DO NOT CHANGE THIS
\usepackage{natbib}  % DO NOT CHANGE THIS AND DO NOT ADD ANY OPTIONS TO IT
\usepackage{caption} % DO NOT CHANGE THIS AND DO NOT ADD ANY OPTIONS TO IT
\usepackage{color}
\usepackage{amsmath,amsfonts}
\usepackage{amssymb}
\newtheorem{assumption}{Assumption}
\usepackage[ruled,vlined,linesnumbered]{algorithm2e} % Load algorithm2e with options
\SetAlgoNlRelativeSize{-1} % Decrease the size of line numbers for clarity
\SetNlSty{bf}{}{} % Bold line numbers with parentheses

\newtheorem{theorem}{Theorem}
\newtheorem{corollary}[theorem]{Corollary}
\allowdisplaybreaks

\nocopyright %-- Your paper will not be published if you use this command
\title{Multi-Agent Path Finding Among Dynamic Uncontrollable Agents\\ with Statistical Safety Guarantees}
\author{
    Kegan J. Strawn\textsuperscript{\rm 1},
    Thomy Phan\textsuperscript{\rm 3},
    Eric Wang\textsuperscript{\rm 1},
    Nora Ayanian\textsuperscript{\rm 2}, 
    Sven Koenig\textsuperscript{\rm 3},
    Lars Lindemann\textsuperscript{\rm 1}\\
}
\affiliations{
    \textsuperscript{\rm 1}University of Southern California\\
    \textsuperscript{\rm 2}Brown University\\
    \textsuperscript{\rm 3}University of California, Irvine\\
    \{k.j.strawn, llindema\}@usc.edu,
    nora\_ayanian@brown.edu,
    \{thomyp, sven.koenig\}@uci.edu
}
\begin{document}

\maketitle

\begin{abstract}
Existing multi-agent path finding (MAPF) solvers do not account for uncertain behavior of uncontrollable agents. We present a novel variant of Enhanced Conflict-Based Search (ECBS), for both one-shot and lifelong MAPF in dynamic environments with uncontrollable agents. Our method consists of (1) training a learned predictor for the movement of uncontrollable agents, (2) quantifying the prediction error using conformal prediction (CP), a tool for statistical uncertainty quantification, and (3) integrating these uncertainty intervals into our modified ECBS solver. Our method can account for uncertain agent behavior, comes with statistical guarantees on collision-free paths for one-shot missions, and scales to lifelong missions with a receding horizon sequence of one-shot instances. We run our algorithm, CP-Solver, across warehouse and game maps, with competitive throughput and reduced collisions.
\end{abstract}


\section{Introduction}
\begin{figure}[t]
\centering
\includegraphics[width=0.80\columnwidth]{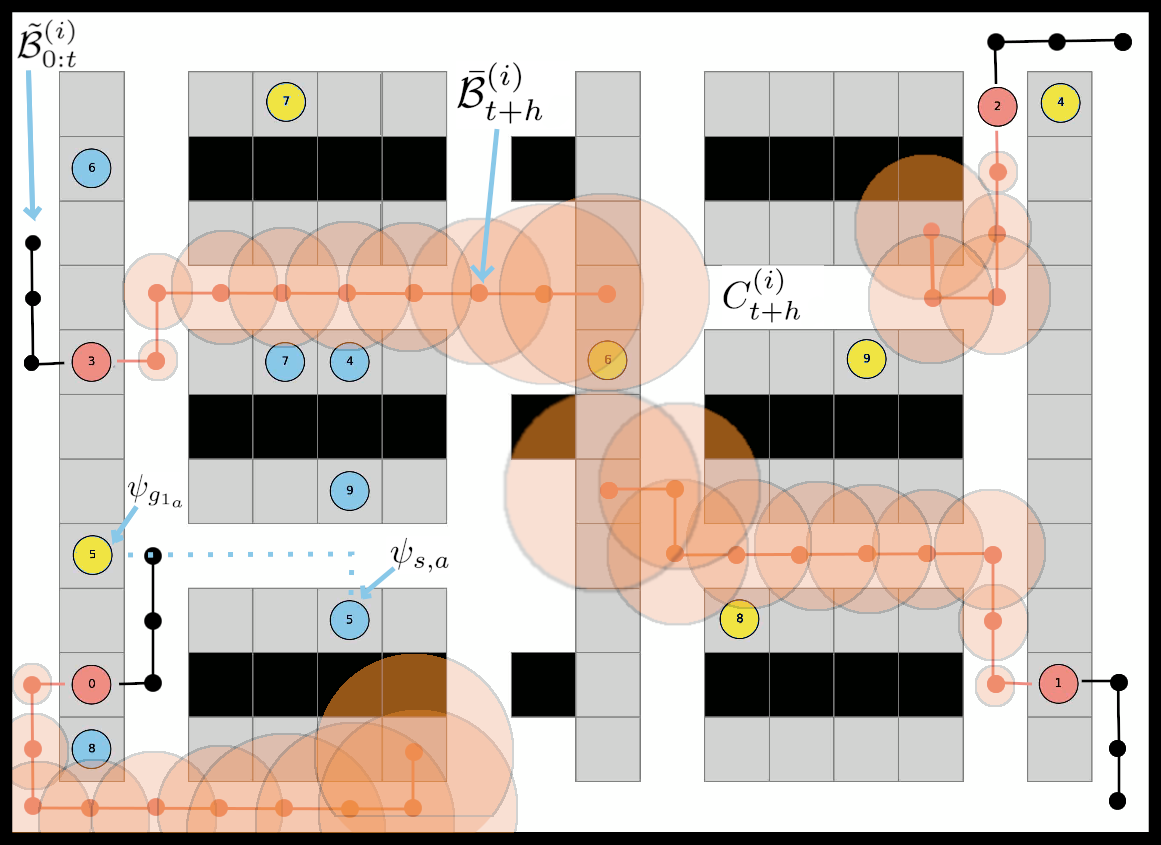}
\caption{CP-Solver plans paths across a mission interval $[t, T]$ for controlled agents $a\in A := \{4, 5, 6, 7, 8, 9\}$ from starting locations $\psi_{s, a}$ (blue circles) to goal locations $\psi_{g,a}$ (yellow circles) such that they avoid uncontrollable agents $b\in U := \{0, 1, 2, 3\}$ (solid red circles) with probability $1-\delta$. This is achieved via predictions $\bar{\mathcal{B}}_{t+1:t+H}^{(i)}$ (red lines) and uncertainty intervals $C_{t+1:t+H} $ (red shaded circles) computed from observations $\tilde{\mathcal{B}}_{0:t}^{(i)}$.}
\label{examplefig}
\end{figure}

\begin{figure}[t]
\centering
\includegraphics[width=\columnwidth]{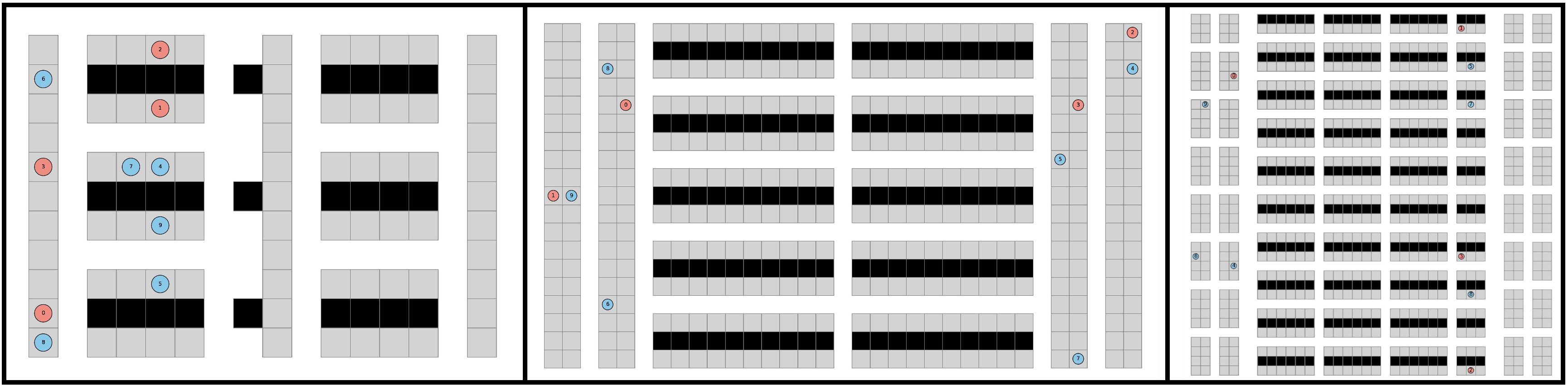}
\caption{Three MAPF DUA warehouse instances (small, medium, large) with 6 controlled agents (blue) and 4 dynamic agents (red).}
\label{warehouse_maps}
\end{figure}

Multi-Agent Path Finding (MAPF), where multiple agents must navigate to goal locations without collisions, has broad applications in robotics, video games, and logistics~\cite{lavalle2006planning, yu2016optimal, stern2019mapf}. While  MAPF algorithms are effective in one-shot (start-to-goal) static environments, they do not handle settings where agents share the environment with uncontrollable, non-cooperative agents, such as humans or player-controlled agents in a game. This paper focuses on MAPF for controllable agents that interact with these uncontrollable agents whose plans and behaviors are unknown. We provide an approach applicable to  the one-shot and lifelong variants of MAPF where, in lifelong MAPF, the agents attend to a potentially endless sequence of goals~\cite{ma2017lifelong}. Specifically:
\begin{itemize}
\item We formulate the problem of MAPF  among Dynamic Uncontrollable Agents (DUA).
\item We propose CP-Solver which combines a (learned) predictive model of uncontrollable agents and conformal prediction (CP)  with Enhanced Conflict-Based Search (ECBS). CP-solver equips predictions with uncertainty intervals and prioritizes predictions during ECBS conflict resolution, see Figure~\ref{examplefig}.
\item We present two variants of CP-Solver: open-loop for one-shot DUA and closed-loop for lifelong DUA.
\item We test multiple benchmarks (see Figures~\ref{warehouse_maps} and~\ref{videogamemaps}), demonstrating competitive throughput and runtime with statistical guarantees on collision avoidance.
\end{itemize}

\section{Related Work}
Traditional MAPF is focused on static environments in a one-shot setting producing a plan for all agents before executing any actions~\cite{stern2019mapf}. Finding a solution of collision-free paths for all agents is an NP-hard problem, yet popular solvers such as Conflict-Based Search (CBS) and its extensions scale in the number of conflicts rather than agents~\cite{sharon2015conflict,barer2014suboptimal}. Recent work in MAPF has shifted to address issues of scalability, robustness to delays or failures, and lifelong scenarios~\cite{stern2019mapf,strawn2022byzantine,bellusci2020multi,ma2017lifelong,wan2018lifelong}. In lifelong MAPF agents attend to a sequence of goals over an indefinite time horizon, planning as they take actions. Most solutions split the lifelong problem into multiple MAPF instances and iteratively resolve conflicts~\cite{li2021lifelong,ma2017lifelong}. 
 
Existing MAPF strategies can incorporate real-time updates, allowing controlled agents to react to new information about obstacles or agent locations. Closest to our work is probabilistic MAPF, where delays and agent failures are accounted for with stochastic models and robust plans that are generated to account for  uncertainty of controlled agents~\cite{atzmon2018robust,atzmon2020probabilistic,okumura2023fault,chen2021symmetry,phan2024adaptive}. However, these are computationally expensive, lack statistical safety guarantees, have limited time horizons, focus on agent failures or changing environments, and do not consider planning around dynamic uncontrollable agents in the environment.

Model Predictive Control (MPC) applies actions in a receding horizon, selecting a minimum-cost action sequence repeatedly using predictions of dynamic agents conditioned on the current state and a history of states seen so far to improve motion planning and control~\cite{rawlings2000tutorial}. However, models of uncontrollable agents are usually not available or exactly known. Inspired by predictive control algorithms and recent work in uncertainty quantification of trajectory predictors with conformal prediction (CP), see e.g., ~\cite{angelopoulos2024theoretical,lindemann2023safe,Stankeviciute2021Conformal}, we  develop CP-solver as an extension of ECBS to address MAPF among uncontrollable agents. Alternative approaches model the underlying uncertainty as a Gaussian distribution, use Kalman filters, or use reachability analysis~\cite{Berkenkamp2015Safe,Thrun2005Probabilistic,Ames2021Control,Rober2023Backward,bansal2017hamilton}. These approaches can ensure safety, but are often overly conservative, computationally expensive, or make unrealistic assumptions. 

\section{Preliminaries: Multi-Agent Path Finding}
Let $A := \{1, 2, \ldots, n\}$ be a set of $n$ controllable agents moving on a given undirected and finite graph $G := (V, E)$, where $V \subseteq \mathbb{R}^{N}$ corresponds to a set of reachable vertices, such as locations in an $N := 2$ dimensional grid, and $(\mathcal{B}_{t, a}, \mathcal{B}_{t+1, a}) \in E \subseteq  \mathbb{R}^N \times \mathbb{R}^N$ are edges connecting two adjacent vertices traversed by an agent $a \in A$ at time $t\in [0,T-1]$ where $T>0$ is a final mission horizon. Each edge has a non-negative cost for traversing from $\mathcal{B}_{t,a}$ to $\mathcal{B}_{t+1,a}$, denoted by a given $cost((\mathcal{B}_{t,a}, \mathcal{B}_{t+1, a})) \in [1, \infty)$. All agents, $a \in A$, are assigned (random or user-specified) starting locations $\psi_{s, a} \in V$ and goal locations $\psi_{g, a} \in V$, written as an agent's assignment $\psi_{a} := (\psi_{s, a}, \psi_{g, a})$. Agents select edges to move to a neighboring vertex or wait at the current vertex. Agents wait at their goal until all goals are achieved. All vertices hence have an edge back to themselves. To move between vertices, an agent takes an action at timestep $t$, moving from vertex $\mathcal{B}_{t, a} \in V$ with $\pi_{a} : %\mathbb{R}^{N} \rightarrow \mathbb{R}^{N}
V \times T \times A \rightarrow V$ to their next vertex $\mathcal{B}_{t+1, a} := \pi_{a}(\mathcal{B}_{t, a},t, a)$. To simplify notation, we will use $\pi_{a}(\mathcal{B}_{t, a})$ and thereby suppress the input arguments of the timestep $t$ and the agent $a$. The solver has access to the graph, agents, and timestep to generate a sequence of actions $\pi_{0:T-1, a} := \{\pi_{a}(\mathcal{B}_{0, a}), \ldots, \pi_{a}(\mathcal{B}_{T-1, a})\}$ that produce a sequence of vertices $\mathcal{B}_{0:T, a} := \{\mathcal{B}_{0, a}, \mathcal{B}_{1, a}, \ldots, \mathcal{B}_{T, a}\}$, where $\mathcal{B}_{0, a} := \psi_{s, a}$, $\mathcal{B}_{T, a} := \psi_{g, a}$, $\mathcal{B}_{t,a} \in V, \forall t\in[0,T-1]$ and $(\mathcal{B}_{t,a},\mathcal{B}_{t+1,a})\in E, \forall t\in[0,T-1]$.

A collision at timestep $t$, denoted by $\mathcal{K}_{(t, a, b)} := 1$,  exists if a vertex conflict, denoted by ${\mathcal{K}}_{t, (\mathcal{B}_{t, a}, \mathcal{B}_{t, b})}$, or edge conflict, denoted by $\mathcal{K}_{(\mathcal{B}_{t, a}, \mathcal{B}_{t+1, a}), (\mathcal{B}_{t, b}, \mathcal{B}_{t+1, b})}$, exists where two agents $a, b \in A, a \neq b$ occupy the same location at the same time, i.e., $\mathcal{B}_{t, a} = \mathcal{B}_{t, b}$, or traverse the same edge in opposite directions, i.e., $(\mathcal{B}_{t, a}, \mathcal{B}_{t+1, a}) = (\mathcal{B}_{t+1, b}, \mathcal{B}_{t, b})$, respectively. Two trajectories are conflict-free if for all timesteps they contain no conflicts: $\mathcal{K}_{(t, a, b)} := 0, \forall t \in [0, T]$. 

A solver is {valid} if the joint solution $\pi_{0:T-1}$ contains action sequences $\pi_{0:T-1, a}$ that produce a conflict-free trajectory from start to goal, $\mathcal{B}_{0:T, a}$, for all agents $a \in A$. A solution $\pi_{0:T-1}$ is {optimal} if it is valid and minimizes a pre-defined cost:
\begin{equation}\label{eq:optimal}
    \min_{\pi_{0:T-1}} \sum^{n}_{a=1} cost(\pi_{a}(\mathcal{B}_{t, a})) \forall t \in [0,T-1]
\end{equation}
where $cost(\pi_{a}(\mathcal{B}_{t, a})) := cost((\mathcal{B}_{t, a}, \mathcal{B}_{t+1, a}))$. The service time, denoted by $S(\mathcal{B}_{0:T, a}) := |\mathcal{B}_{0:T, a}|$, is the number of edges in the trajectory. If all edge costs are equal to one, then $\sum_{t=0}^{T-1}cost(\pi_{a}(\mathcal{B}_{t, a})) = S(\mathcal{B}_{0:T, a})$. We note that this is not necessarily equivalent to the makespan, which is the time $T$ when all agents have reached their goals. Solutions are often evaluated by their makespan, service time, and runtime (the time it takes to compute the solution in seconds).

Conflict-Based Search (CBS), a popular MAPF algorithm, finds a valid conflict-free solution $\pi_{0:T-1}$~\cite{sharon2015conflict} by splitting the problem into a high- and low-level search. Using A* to find the shortest path between a start and goal state at the low-level~\cite{hart1968formal} and building a constraint tree at the high level. A new node on the tree is added for the first conflict found, represented as $(a, \mathcal{B}_{t, a}, t)$, forcing the agent $a$ to avoid the vertex $\mathcal{B}_{t, a}$ at timestep $t$. CBS selects the lowest cost solution seen so far, until a solution with no conflicts is found. This guarantees an optimal solution in terms of makespan. However, the search grows exponentially and becomes intractable with runtime limits~\cite{sharon2015conflict}. ECBS is a bounded suboptimal variant based on a focal search, inflating the low-level search heuristic by $w$ while maintaining an additional focal search priority queue at the high-level. The bound $w$ determines how much the resulting solution cost may vary from the optimal solution, speeding up the search significantly~\cite{barer2014suboptimal}. 

\section{Problem Formulation: MAPF DUA}
We define the Multi-Agent Path Finding (MAPF) among Dynamic Uncontrolled Agents (DUA) problem as desiring a set of collision-free action sequences for the set of controllable agents from start to goal locations among a set of dynamic and uncontrollable agents. Therefore, let $U := \{1, 2, \ldots, m\}$ be a set of $m$ uncontrolled dynamic agents. %whose goals and paths are not known in advance and that the MAPF solver does not control. 
All agents in $A$ and $U$ have a starting location and a goal location, denoted by $\psi_{a}, \forall a \in A$ and $\tilde{\psi}_{b}, \forall b \in U$. We seek a valid sequence of actions $\pi_{a}, \forall a \in A$ such that the solver generates paths for each controlled agent from $\psi_{s, a}$ to $\psi_{g, a}$ while minimizing service time and avoiding collisions ($\mathcal{K}_{(t, a, b)} \neq 1$), with uncontrolled agents controlled by an unknown action sequence $\tilde{\pi}$. We write this \textbf{one-shot, open-loop} problem as an ideal optimization: 
\begin{subequations}
    \begin{align}
        \min_{\pi_{0:T-1}} &\sum^{n}_{a=1} (S(\mathcal{B}_{a}))\label{open_a}\\
        \text{subject to} \nonumber \\
        \mathcal{B}_{0, a} &:= \psi_{s, a}, \forall a \in A \label{open_c}\\
        \mathcal{B}_{t+1, a} &:= \pi_{a}(\mathcal{B}_{t, a}), \forall a \in A \label{open_d}\\
        \mathcal{B}_{1:\hat{T}, a} &:= \{\mathcal{B}_{1, a}, \ldots, \mathcal{B}_{T-1, a}, \psi_{g, a}\}, \forall a \in A \label{open_e}\\
        \tilde{\mathcal{B}}_{0, b} &:= \tilde{\psi}_{s, b}, \forall b \in U \label{open_f}\\
        \tilde{\mathcal{B}}_{t+1, b} &:= \tilde{\pi}_{b}(\tilde{\mathcal{B}}_{t, b}), \forall b \in U \label{open_g}\\
        \tilde{\mathcal{B}}_{1:\hat{T}, b} &:= \{\tilde{\mathcal{B}}_{1, b}, \ldots, \tilde{\mathcal{B}}_{T-1, b}, \tilde{\psi}_{g, b}\}, \forall b \in U \label{open_h}\\
        \mathcal{K}_{(t, a, k)} &\neq 1, \forall t \in [0, T], \forall (a, k) \in A \times A \label{open_i}\\
        \mathcal{K}_{(t, a, b)} &\neq 1, \forall t \in [0, T],  \forall a \in A, \forall b \in U \label{open_j}
    \end{align}
\end{subequations}
Solving this problem requires knowledge of the assignment $(\tilde{\psi}_{s, b},\tilde{\psi}_{g, b})$ of uncontrolled agents and their action sequence $\tilde{\pi}_{t,b}$, which are not available. In general, the open-loop (one-shot) problem becomes intractable as the graph size, agent set, and minimum makespan increase~\cite{stern2019mapf}. 

In lifelong MAPF all agents in $A$ and $U$ have a starting location and a $q_{a} > 0$ length sequence of locations to attend to sequentially, $\psi_{g_{a}} := \{\psi_{g_{{1}_{a}}}, \psi_{g_{{2}_{a}}}, \ldots, \psi_{g_{q_{a}}}\}, \forall a \in A$ and $\tilde{\psi}_{g_{b}} := \{\tilde{\psi}_{g_{{1}_{b}}}, \tilde{\psi}_{g_{{1}_{b}}}, \ldots, \tilde{\psi}_{g_{q_{b}}, b}\}, \forall b \in U$. In the closed-loop (lifelong) DUA problem, we change the minimization of service time to maximization of throughput: $\frac{1}{\hat{T}}\Gamma(\mathcal{B}_{0:\hat{T}, a})$ where $\Gamma(\mathcal{B}_{0:\hat{T}, a})$ is the number of goals a trajectory achieves over a user-set long time horizon $\hat{T} > 0$. We write this \textbf{lifelong, closed-loop} problem (L-MAPF DUA) as an ideal optimization: 
\begin{subequations}
    \begin{align}
        \underset{\pi_{0:\hat{T}-1}}{\text{max}} &\frac{1}{\hat{T}}\sum^{n}_{a=1} (\Gamma(\mathcal{B}_{a}))  \label{closed_a}\\
        \text{subject to} \nonumber\\
        \mathcal{B}_{0, a} &:= \psi_{s, a}, \forall a \in A \label{closed_c}\\
        \mathcal{B}_{t+1, a} &:= \pi_{a}(\mathcal{B}_{t, a}), \forall a \in A \label{closed_d}\\
        \mathcal{B}_{1:\hat{T}, a} &:= \{\mathcal{B}_{1, {a}}, \ldots, \psi_{g_{1_{a}}}, \ldots, \mathcal{B}_{\hat{T}-1, a}, \psi_{g_{q_{a}}}\}, \forall a \in A \label{closed_e}\\
        \tilde{\mathcal{B}}_{0, b} &:= \tilde{\psi}_{s, b}, \forall b \in U \label{closed_f}\\
        \tilde{\mathcal{B}}_{t+1, b} &:= \tilde{\pi}_{b}(\tilde{\mathcal{B}}_{t, b}), \forall b \in U \label{closed_g}\\
        \tilde{\mathcal{B}}_{1:\hat{T}, b} &:= \{\tilde{\mathcal{B}}_{1, b}, \ldots, \tilde{\psi}_{g_{{1}_{b}}}, \ldots, \tilde{\mathcal{B}}_{\hat{T}-1, b}, \tilde{\psi}_{g_{q_{b}}}\}, \forall b \in U \label{closed_h}\\
        \mathcal{K}_{(t, a, k)} &\neq 1, \forall t \in [0, \hat{T}], \forall (a, k) \in A \times A \label{closed_i}\\
        \mathcal{K}_{(t, a, b)} &\neq 1, \forall t \in [0, \hat{T}],  \forall a \in A, \forall b \in U \label{closed_j}
    \end{align}
\end{subequations}

\section{Approach: CP-Solver}
The uncontrolled agents' action sequences and assignments in equations \eqref{open_f}-\eqref{open_h} and \eqref{closed_f}-\eqref{closed_h} are typically unknown, e.g., in the case of a human agent. Thus, it is impossible to solve the DUA problem exactly. We instead focus on an approximate solution where we aim for probabilistic collision avoidance by replacing equations \eqref{open_j} and \eqref{closed_j} with:
\begin{equation}
    \begin{aligned}
        \text{Prob}(\mathcal{K}_{(t, a, b)} \neq 1, \forall t \in [0, \bar{T}],  \forall a \in A, \forall b \in U) \geq 1-\delta
    \end{aligned}
\end{equation}
where $\bar{T}\in \{T,\hat{T}\}$ and $\delta \in (0, 1)$ is a user defined failure probability. To solve this approximate problem, we advocate for an approach that (1) predicts the paths of dynamic agents, and (2) quantifies prediction uncertainty statistically. 

%%%%%%%%%%%%%%%% Trajectory Prediction %%%%%%%%%%%%%%%%%%%%%%%%%
\subsection{Trajectory Prediction} 
Let $\mathcal{D}_{\tilde{B}}$ be an unknown distribution over uncontrolled dynamic agent trajectories. In this case, the random trajectory $(\tilde{\mathcal{B}}_{0}, \tilde{\mathcal{B}}_{1}, \cdots) \sim \mathcal{D}_{\tilde{B}}$ is generated by $\tilde{\mathcal{B}}_{t+1} := \tilde{\pi}(\tilde{\mathcal{B}}_{t})$  where the stacked agent states $\tilde{\mathcal{B}}_{t} := (\tilde{\mathcal{B}}_{t, 1}, \cdots , \tilde{\mathcal{B}}_{t, m})$ at time $t$ are drawn from $\mathbb{R}^{Nm}$. We make no assumptions on the form of the distribution $\mathcal{D}_{\tilde{B}}$ but assume 1) the availability of data independently drawn from $\mathcal{D}_{\tilde{B}}$ generated by $\tilde{\mathcal{B}}_{t+1} := \tilde{\pi}(\tilde{\mathcal{B}}_{t})$, and 2) that $\mathcal{D}_{\tilde{B}}$ is independent of any controlled agents. 

\begin{assumption}  \label{data_assumption}
We have a dataset of trajectories $D := \{\tilde{\mathcal{B}}^{(1)}, \cdots, \tilde{\mathcal{B}}^{(d)}\}$ in which each of the $d$ trajectories $\tilde{\mathcal{B}}^{(i)} := \{\tilde{\mathcal{B}}_{0}^{(i)} , \tilde{\mathcal{B}}_{1}^{(i)}, \cdots\}$ is independently drawn from $\mathcal{D}_{\tilde{B}}$.
\end{assumption}

Assumption~\ref{data_assumption} is generally not restrictive, especially in warehouse environments, as  data can be recorded before deployment or obtained from rapidly advancing high-fidelity simulators and open datasets~\cite{padalkar2023open}. We split the  dataset of trajectories $D$ into separate training datasets $D_{train}$, $D_{val}$, and $D_{test}$ from which we will train, validate, and test a trajectory predictor. Additionally, setting aside a small dataset $D_{cal}$ to quantify prediction uncertainty.

\begin{assumption} \label{interference_assumption}
For any time $t \geq 0$, the controlled agent sequences $(\pi_{a}(\mathcal{B}_{0, a}),\cdots,\pi_{a}(\mathcal{B}_{t-1, a})), \forall a \in A$ and the resulting trajectories $(\mathcal{B}_{0},\cdots,\mathcal{B}_{t})$, do not change the distribution of dynamic agent trajectories $(\tilde{\mathcal{B}}_{0}, \tilde{\mathcal{B}}_{1}, \cdots) \sim \mathcal{D}_{\tilde{B}}$.
\end{assumption}

Assumption~\ref{interference_assumption} holds approximately in many applications, e.g., when controlled agents operate in socially acceptable ways without changing the intentions of other agents. In our experiments, the dynamic uncontrollable agents are given the right of way and are not influenced by controlled agents. Furthermore, it has been shown that CP guarantees remain valid even under small distribution shifts~\cite{cauchois2023robust,strawn2023conformal}. In cases where such interaction is present, robust uncertainty quantification could help preserve guarantees~\cite{angelopoulos2024theoretical,cauchois2023robust}. 
Given a prediction interval $\lambda :=[1, H]$, where $H \in [0, \infty)$ is a prediction horizon, and the history of dynamic agent observations $\tilde{\mathcal{B}}_{0:t}$, we seek a trajectory predictor $Y: \mathbb{R}^{(t+1)Nm} \rightarrow \mathbb{R}^{NmH}$ that predicts the $H$ future agent vertices $(\tilde{\mathcal{B}}_{t+1},\hdots,\tilde{\mathcal{B}}_{t+H})$ as $\bar{\mathcal{B}}_{\lambda} := Y(\tilde{\mathcal{B}}_{0:t})$ where $\bar{\mathcal{B}}_{\lambda} := (\bar{\mathcal{B}}_{t+1}, \cdots, \bar{\mathcal{B}}_{t+H})$.
In principle, we can use any trajectory predictor $Y$, e.g.,  long short-term memory networks~\cite{alahi2016social,hochreiter1997long} or transformer architectures~\cite{nayakanti2022wayformer}. We independently sample the training dataset $D_{train}$ from $\mathcal{D}_{\tilde{B}}$ with trajectories of length ${T}$, where  $\mathcal{\tilde{B}}^{(i)}_{0:{T}} := (\mathcal{\tilde{B}}_{0}^{(i)}, \cdots, \mathcal{\tilde{B}}_{t}^{(i)}, \mathcal{\tilde{B}}_{t+1}^{(i)}, \cdots,  \mathcal{\tilde{B}}_{{T}}^{(i)})$ is the $i$-th trajectory in the dataset. We train the predictor  by minimizing the following loss function over $D_{train}$, validating with $D_{val}$ and testing accuracy with $D_{test}$:  
\begin{equation}\label{eq:train_pred}
    \begin{aligned}
    \min_{Y} \frac{1}{|D_{train}|}\sum_{i=1}^{|D_{train}|}\|\mathcal{\tilde{B}}^{(i)}_{\lambda} - Y(\tilde{\mathcal{B}}_{0:t}^{(i)})\|^2.
    \end{aligned}
\end{equation} 
\subsection{Uncertainty Quantification of Predictions}\label{prelim:CP}
We use conformal prediction (CP) to construct statistical regions around the predicted trajectories that contain the true, yet unknown, agent trajectory with a user-defined confidence level~\cite{Angelopoulos2021Gentle,lindemann2024formal}. The CP method we use constructs valid prediction regions for any learned time series predictor $Y$. 

In Algorithm~\ref{alg:cpAlg}, we present a modified version of a recent framework from \cite{cleaveland2024conformal} that uses linear complementarity programming (LCP) to obtain tight trajectory prediction regions by compensating for prediction errors across timesteps. In contrast to \cite{cleaveland2024conformal}, we incorporate reasoning over multiple uncontrollable agents. In LCP CP, normalization constants $\alpha_{t+h}$ are introduced to normalize prediction errors at each time step, resulting in less conservative prediction regions compared to other existing work. We summarize  Algorithm~\ref{alg:cpAlg} next.

Given a failure probability $\delta \in (0, 1)$, a history of observations $\tilde{\mathcal{B}}_{0:t} := (\tilde{\mathcal{B}}_{0}, \cdots, \tilde{\mathcal{B}}_{t})$ at time $t$ for all $m$ dynamic agents $\tilde{\mathcal{B}}_{t} := (\tilde{\mathcal{B}}_{t, 0}, \cdots, \tilde{\mathcal{B}}_{t, m})$, and trajectory predictor $Y$ producing predictions $\bar{\mathcal{B}}_{\lambda} := (\bar{\mathcal{B}}_{t+1}, \cdots, \bar{\mathcal{B}}_{t+H})$ for the specified prediction horizon $\lambda$, Algorithm~\ref{alg:cpAlg} generates values ${C}_{\lambda} := ({C}_{t+1}, \cdots, {C}_{t+H})$ as prediction intervals around each prediction that guarantee:
\begin{equation}\label{eq:prob_gua}
    \text{Prob}(R(\bar{\mathcal{B}}_{t+h, b}, \tilde{\mathcal{B}}_{t+h,b}) \leq {C}_{t+h}, \text{ }\forall b \in U, \forall h \in \lambda) \geq 1 - \delta
\end{equation}
where $R(\bar{\mathcal{B}}_{t+h}, \tilde{\mathcal{B}}_{t+h})$ is the prediction error:
\begin{equation}\label{eq:nonconformity_function}
    R(\bar{\mathcal{B}}_{t+h}, \tilde{\mathcal{B}}_{t+h}) := \parallel\bar{\mathcal{B}}_{t+h} - \tilde{\mathcal{B}}_{t+h}\parallel.
\end{equation}
We start the CP process by first computing the max prediction error $\bar{{C}}_{t+h}^{(i)} := \underset{b \in U}{\max} (R(\bar{\mathcal{B}}_{t+h, b}^{(i)}, \tilde{\mathcal{B}}_{t+h, b}^{(i)}))$ at each timestep across instances in the calibration dataset $D_{cal}$ (lines~\ref{predictionErrorLine}-\ref{predictionErrorLineEnd}). Then, we split the prediction errors $\bar{C}_{\lambda}$ into $C_{\lambda, cal1}$ and $C_{\lambda, cal2}$. Next, we follow two key steps: 
\begin{enumerate}
    \item Computing normalization constants $\alpha_{t+h} \geq 0, \forall h \in \lambda$ with $C_{\lambda, cal1}$ according to \cite{cleaveland2024conformal}, referred to as the routine LCP$(\cdot)$ (line~\ref{LPOpt}). 
    \item Taking the maximum across timesteps of the values $\hat{C}^{(i)}_{cal2} := \max(\alpha_{t+1}{C}_{t+1, cal2}^{(i)}, \ldots, \alpha_{t+H}{C}_{t+H, cal2}^{(i)})$ and sorting them in non-decreasing order (lines~\ref{maxAndSort}-\ref{maxAndSortEnd}). 
\end{enumerate}
Afterwards, we append infinity as the $(|D_{cal, 2}| + 1)$-th value (line~\ref{appendInfty}) before setting $C_{t+h}$ as the  $p := \lceil (|D_{cal, 2}| + 1)(1 - \delta)\rceil$-th smallest value of $\hat{C}^{(i)}_{cal2}$ divided by $\alpha_{t+h}$ (lines~\ref{setp}-\ref{intervalPrep}). The produced CP intervals $C_{t+h}$ are guaranteed to satisfy \eqref{eq:prob_gua} and are later passed to Algorithm~\ref{alg:generalAlg} within CP-Solver.

\begin{algorithm}
\caption{Conformal Prediction Setup}\label{alg:cpAlg}
\KwIn{Datasets $(D_{train}, D_{val}, D_{test}, D_{cal})$, Confidence $\delta$, Horizon $\lambda$, Timesteps $(t, T)$}
\KwOut{Predictor $Y$, CP Intervals $C_{\lambda}$}
$\text{Learn Predictor } Y \text{ from } D_{train}, D_{val}, D_{test} \text{ (Eq. \ref{eq:train_pred})}$\;
    \For{$i$ in $D_{cal}$} { \label{predictionErrorLine}
    $\bar{\mathcal{B}}_{\lambda}^{(i)} \leftarrow Y(\tilde{\mathcal{B}}_{0:t}^{(i)})$\;
        \For{$h \in [1, H]$}{
            $\text{ Compute } \bar{{C}}_{t+h}^{(i)} \leftarrow \underset{b \in U}{\max} (R(\bar{\mathcal{B}}_{t+h, b}^{(i)}, \tilde{\mathcal{B}}_{t+h, b}^{(i)}))$\;
        }
    }\label{predictionErrorLineEnd}
    $\bar{C}_{\lambda, cal1}, \bar{C}_{\lambda, cal2} \leftarrow \text{Split } \bar{C}_{\lambda}$\;
    $\{\alpha_{t+1}, \ldots,\alpha_{t+H}\} \leftarrow \text{LCP}(\bar{C}_{\lambda, cal1}, \delta)$\;\label{LPOpt}
    \For{$i$ in $\bar{C}_{\lambda, cal2}$}{\label{maxAndSort}
    $\hat{C}_{cal2}^{(i)} \leftarrow \max( \alpha_{t+1}\bar{C}_{t+1, cal2}^{(i)}, \ldots, \alpha_{t+H}\bar{C}_{t+H, cal2}^{(i)})$\;
}
$\text{Sort } \hat{C}_{cal2} \text{ in non-decreasing order}$\;\label{sortCP}\label{maxAndSortEnd}
$\text{Append } \hat{C}^{(|D_{cal,2}|+1)}_{cal2} \leftarrow \infty$\;\label{appendInfty}
$\text{Set } p \leftarrow \lceil(|D_{cal, 2}| + 1)(1-\delta))\rceil$\;\label{setp}
\For{$h \in \lambda$}{ 
    $\text{Set } C_{t+h} \leftarrow \frac{\hat{C}^{(p)}_{cal2}}{\alpha_{t+h}}$\;\label{intervalPrep}
}
%}
\end{algorithm}
%%%%%%%%%%%%%%%%%%%%%%%% CP %%%%%%%%%%%%%%%%%%%%%%%%%%%%%%%%%%%

\subsection{Open-Loop CP-Solver}
In Algorithm~\ref{alg:generalAlg}, we present our Conformal Predictive Solver (CP-Solver). Before planning a path for the controlled robots, we run Algorithm~\ref{alg:cpAlg} to obtain CP intervals $C_{\lambda} := \{C_{\lambda, 1}, \ldots, C_{\lambda, m}\}$. During planning, i.e., at runtime $t$, we have access to the observations $\tilde{\mathcal{B}}_{0:t}$ of dynamic agents from which we compute predictions $\bar{\mathcal{B}}_{\lambda} := Y(\tilde{\mathcal{B}}_{0:t})$ (lines~\ref{initialObservations}-\ref{computePredictions}). 

We first project our CP intervals satisfying equation~\eqref{eq:prob_gua} onto the discrete graph $G$ by building CP interval sets. This is a crucial step in our method that is needed since the CP intervals define a region in $\mathbb{R}^N$, while MAPF operates on discrete graphs $G$. Our method builds a set of CP interval vertices at every timestep: $\bar{\mathcal{B}}_{\lambda, C} := \{\bar{\mathcal{B}}_{t+1, C}, \ldots, \bar{\mathcal{B}}_{t+H, C}\}$ via the $\text{discretize}(\cdot)$ function (line~\ref{discretize}). Specifically, the set $\bar{\mathcal{B}}_{t+h, C} := \{v_{0}, v_{1}, \ldots, v_{\rho_{t+h}}\}$ contains $\rho_{t+h}$ vertices $v_j\in V$ that satisfy: 1)  $\parallel \bar{\mathcal{B}}_{t+h, b} - v_j\parallel \leq C_{t+h}$ for some agent $b\in U$, i.e., $v_j$ is $C_{t+h}$-close to the predictions of an uncontrollable agent $b$, and 2) the shortest path is $SP(\tilde{\mathcal{B}}_{t,b}, v_j)\le h$ so that $v_j$ can be reached from $\tilde{\mathcal{B}}_{t,b}$ in less than $h$ edges. If the controllable agents in $A$ avoid the CP interval vertex sets $\bar{\mathcal{B}}_{\lambda, C}$ on $G$, it follows that collisions with uncontrollable agents are avoided with a probability of no less than $1 -\delta$. 

%%%%%%%%%%%%%%%%%%%%%%%%%%%%%%%%%%%%%%%%%%%%%%%%%%%%%%%%%%%%%%%%%
We begin our modified ECBS algorithm with an initial node $Z$ and an empty set of constraints (line~\ref{initConstraints}). The high-level node stores: the constraints on trajectories for each controlled agent, predicted trajectories of uncontrolled agents, discretized CP interval sets, and controlled agent paths. These controlled agent paths are found with A* in the $\text{low\_level}(\cdot)$ search and stored as action sequences into $Z_{\pi_{t:T-1}}$ (lines~\ref{lowlevel}-\ref{controlled_path}). The cost of the trajectories, $Z_{cost}$, is the service time for each controlled agent trajectory (line~\ref{cost}). The node $Z$ is added to a minimum cost-based priority queue, which our modified ECBS solver searches over (line~\ref{priorityqueue}). 

%%%%%%%%%%%%%%%%%% ALGORITHM 2 %%%%%%%%%%%%%%%%%%%%%%%%%%%%%%%%
\begin{algorithm}
\caption{Open-Loop CP-Solver}\label{alg:generalAlg}
\KwIn{Graph $G$, Agents $A$, Dynamic Agents $U$, Assignments $\psi$, Horizon $\lambda$, Trained Predictor $Y$, Conformal Intervals $C_{\lambda}$, timestep $t$}
\KwOut{Solution $Z_{\pi_{t:T-1}}$ and Trajectories $\mathcal{B}_{t+1:t+H}$}
%\While{${\mathcal{B}}_{t, k} \neq {\psi}_{g, k}, \forall k \in {A}$}{
$\text{Set }\tilde{\mathcal{B}}_{0:t} \leftarrow observe(U, t)$\;\label{initialObservations}
$\text{Set }\bar{\mathcal{B}}_{\lambda} \leftarrow Y(\tilde{\mathcal{B}}_{0:t})$\;\label{computePredictions}
$\text{Set } \bar{\mathcal{B}}_{\lambda, C} \leftarrow \text{discretize}(C_{\lambda}, \bar{\mathcal{B}}_{\lambda})$\label{discretize}
    
$\text{Set }Z_{constraints} \leftarrow \emptyset$\;\label{initConstraints}
$\textcolor{black}{\text{Set }\mathcal{B}_{0:T}} \leftarrow \text{low\_level}(Z_{\text{constraints}}, G, A, \psi)$\;\label{lowlevel}
$\text{Set }Z_{\pi_{t:T-1}} \leftarrow \bar{\mathcal{B}}_{\lambda, C} \cup \bar{\mathcal{B}}_{\lambda} \cup \mathcal{B}_{0:T}$\;\label{controlled_path} 
$\text{Set }Z_{cost} \leftarrow \sum_{i=0}^{n} S(\mathcal{B}_{0:T, i})$\; \label{cost}
$\text{Insert }OPEN \leftarrow Z \text{ into the priority queue}$\;\label{priorityqueue}

\While{OPEN not empty}{
    $\text{Set } Z \leftarrow $ \textcolor{black}{minimum $Z_{cost}$} node from OPEN\; \label{minNode}
    $\text{Get }\textcolor{black}{\mathcal{K}_{t,a,b}} \leftarrow \text{first\_conflict}(Z_{\pi_{t:T-1}})$\; \label{firstConflict}
    
    \If{$\textcolor{black}{\mathcal{K}_{t,a,b}}$ is 0}{
        \textbf{break}\; \label{foundSolution}
    }
    \If{agent $a, b$ in $U$}{\label{twoUncontrolled}
        % Dynamic agents logic can be inserted here
        $\text{Set }\phi \leftarrow \varnothing;$
    }
    \ElseIf{agent $a$ in ${A}$ and $b$ in $U$}{ \label{oneUncontrolledA}
        % Logic for agent a being dynamic can be inserted here
        $\text{Set }\phi \leftarrow (a);$
    }
    \ElseIf{agent $a$ in ${U}$ and $b$ in $A$}{
        % Logic for agent b being dynamic can be inserted here
        $\phi \leftarrow (b);$\label{oneUncontrolledB}
    }
    \Else{
        $\phi \leftarrow (a, b);$
    }
    \For{$r$ in $\phi$} {\label{iterateConstraints}
        $\text{Set }Z_{r} \leftarrow \text{new\_node}(Z)$\;\label{newNode}
        $\text{Set }Z_{\text{constraints}, r} \leftarrow (\mathcal{K}_{t, a, b}, r)$\;
        $\text{Set }\mathcal{B}_{0:T, r} \leftarrow \text{low\_level}(Z_{\text{constraints}, r}, G, r, \psi_{r})$\;
        $\text{Set }Z_{\pi_{t:T-1}, r} \leftarrow \text{update\_solution}(\mathcal{B}_{0:T, r})$\;
        $\text{Set }Z_{cost, r} \leftarrow Z_{cost} - Z_{cost, r} + S(\mathcal{B}_{0:T, r})$\;
        $\text{Insert }OPEN \leftarrow$ $Z_{r}$\;   \label{insertToQueue}
    }
}
$\text{Set }\mathcal{B}_{t+1:t+H} \leftarrow Z_{\pi_{t:T-1}}(\mathcal{B}_{t:T-1})$\;
\end{algorithm}
%%%%%%%%%%%%%%%%%%%%%%%%
We continue by selecting the minimum cost node from  $OPEN$  until $OPEN$ is empty (lines 9-\ref{minNode}). Then, each node $Z$ is checked for a conflict between pairs of agents, predicted paths, and CP interval vertices (line~\ref{firstConflict}). We do not add a constraint to the set of constraints, denoted by $\phi$, if the conflict involves two uncontrolled agents \textcolor{black}{(line \ref{twoUncontrolled})}. If the conflict involves one uncontrolled agent or CP interval vertex, only the controlled agent receives a constraint (lines \ref{oneUncontrolledA}-\ref{oneUncontrolledB}). If the conflict involves both uncontrolled agents, both receive a constraint (line 21). We then iterate over the found constraint's \textcolor{black}{agents (line \ref{iterateConstraints})} to build a new node $Z_{p}$ for each $p \in \phi$ that inherits from the current node $Z$ through $\text{new\_node}(\cdot)$ (line~\ref{newNode}) \textcolor{black}{and add it to the priority queue (line \ref{insertToQueue})}.  If a conflict-free solution is found (line~\ref{foundSolution}), we execute the action sequence $\pi_{t:T-1}$ for all controlled agents, avoiding the CP interval sets $\bar{\mathcal{B}}_{\lambda, C}$. Thus, equation~\eqref{eq:prob_gua} is satisfied, and our algorithm contributes a probabilistically safe solution to the one-shot DUA problem. 
\begin{corollary}\label{cpcorollary}
    Algorithm~\ref{alg:cpAlg} guarantees that equation~\eqref{eq:prob_gua} is satisfied. Then, by construction, a valid solution $\pi_{\textcolor{black}{0:T-1}}$ for controlled agents produced by CP-Solver in Algorithm~\ref{alg:generalAlg}  guarantees that collision as per equations \eqref{open_i}-\eqref{open_j} are avoided with the user-specified confidence level of $1 - \delta$.
\end{corollary}

\subsection{Closed-Loop CP-Solver}
In the open-loop method, the plan is produced and then executed until all agents reach their goals. Next, we adapt CP-Solver to solve lifelong MAPF DUA (L-MAPF DUA) by adopting a rolling-horizon conflict resolution (RHCR) framework~\cite{li2021lifelong}. To do so, we select an ECBS solver conflict horizon, denoted by $\hat{w}$, and a replanning window, denoted by $H$, \textcolor{black}{where $H \leq \hat{w}$. In RHCR, conflicts are resolved up to this maximum timestep $\hat{w}$ and controlled agents replan their paths every $H$ timesteps~\cite{li2021lifelong}. In our implementation, we set the prediction horizon as $\lambda := [t+1, t+H]$ using the replanning horizon $H$}. While effective in practice, we note that RHCR algorithms are in general incomplete. The RHCR framework may overlook long-term conflicts, occasionally deadlocking. We  extend our CP-Solver algorithm by iteratively solving multiple $H$-windowed open-loop instances, updating our observations of the dynamic agents at the end of each windowed planning horizon. We summarize our algorithm next.

Before planning and executing a user-defined $\hat{T}$ total timesteps, we train the predictor and run the CP process as done in Algorithm~\ref{alg:cpAlg}. Our closed-loop approach then tracks the current timestep $t\in \{0, H, 2H, \cdots, \hat{T}]$, observing the states of the dynamic agents every $H$ steps, and stores their locations in the history of observations $\tilde{\mathcal{B}}_{0:t}$.  We then run Algorithm 2 as a windowed instance of the open-loop approach and in a rolling-horizon fashion. Here, observations are input to the predictor to produce predictions $\bar{\mathcal{B}}_{\lambda}$ that are then used to generate discretized CP interval sets $\bar{B}_{\lambda, C}$ for all dynamic agents in $U$ with $\text{discretize}(\cdot)$.  Once a conflict-free solution is found by Algorithm 2, we apply the action sequence $\pi_{a}(\mathcal{B}_{h, a}), \forall a \in A, \forall h\in[t, t+H-1]$, update $t := t+H$, and repeat while $t < \hat{T}$. Agents whose shortest path on $G$ to their current sequence of goals is less than the replanning horizon: $SP(\mathcal{B}_{t, a}, \psi_{a}) \leq H$, are randomly assigned additional goal locations to visit. At the end of each executed window, we observe the uncontrolled agent movements and solve the next $H$-windowed MAPF DUA instance. 

\textbf{Solvability}: Not all MAPF instances (composed of a graph $G$, controlled agents $A$, and task assignments $\psi$) are solvable~\cite{sharon2015conflict}. A sufficient condition for solvability \textcolor{black}{(feasible to produce a solution)} is that the instance is well-formed~\cite{ma2017lifelong}. In traditional MAPF, an instance is well-formed if and only if the number of tasks is finite, there are exactly as many or more task spots \textcolor{black}{(available vertices, $g \in V$, for goals)} as the number of agents, and for any assignment there exists a path between the start and goal vertices that does not cross another goal vertex. However, in lifelong settings, these conditions may not hold with new tasks arriving over an infinite mission horizon. To address this, it is common to introduce \textcolor{black}{assumptions that maintain} the well-formed condition for lifelong MAPF, such as reserving goal vertices~\cite{ma2017lifelong}. In L-MAPF DUA, dynamic uncontrolled agents may break the well-formed assumption, e.g., an uncontrolled agent stops and prevents another vertex from being reached. To preserve solvability, we introduce the following three-part assumption.

\begin{assumption} \label{wellformed_assumption}
\textcolor{black}{First, we reserve the assigned goal vertices, i.e.,} all goal assignments, $\psi_{g,a}$ for all $a\in A$, are in designated task spots that only the assigned agent can enter after moving out of their starting vertex. \textcolor{black}{Second, if} the controlled agent cannot find a path to its goal, the agent can remain where it is and be excluded from the input to our modified ECBS solver. \textcolor{black}{Third, we assume all uncontrolled agents will eventually reach an assigned goal vertex or keep moving. }
\end{assumption}
We maintain the well-formed attribute by allowing a controlled agent to pause and an uncontrolled agent to proceed with the next planning window if needed to prevent a deadlock. We note that these interactions are undesirable and will be counted as collisions during the experiments.  

\textcolor{black}{\textbf{Real-time Applicability}: Real-world L-MAPF DUA with windowed horizons requires computing and executing actions before uncontrolled agents physically move beyond the given window. This is also true for traditional MAPF, which assumes agents' solutions can be computed fast enough for real-world applications.  To ensure real-time applicability, we make the following assumptions. } %Additionally, as in traditional MAPF, we assume that plans can be made in discrete time and later applied to real robots with continuous control.  
First, all agents begin and complete their action sequences during the execution window $[t, t+H-1]$. At each transition, observations of the dynamic agents are obtained, $\tilde{\mathcal{B}}_{0:t}$, and solutions $\pi_{t,t+H-1}$ for the next horizon are produced before the following execution window begins. Second, uncontrolled agents do not move across more than one edge during a timestep transition $[t, t+1]$. ECBS and RHCR were selected to prioritize speed in re-planning and uphold these assumptions. Future work could explore ECBS alternatives, calibrate the time horizons to decrease runtime further, or integrate CP-Solver with continuous-space forms of MAPF~\cite{andreychuk2022multi,honig2016multi}.

\section{Experimental Evaluation}
\begin{figure}[t]
\centering
\includegraphics[width=0.50\columnwidth]{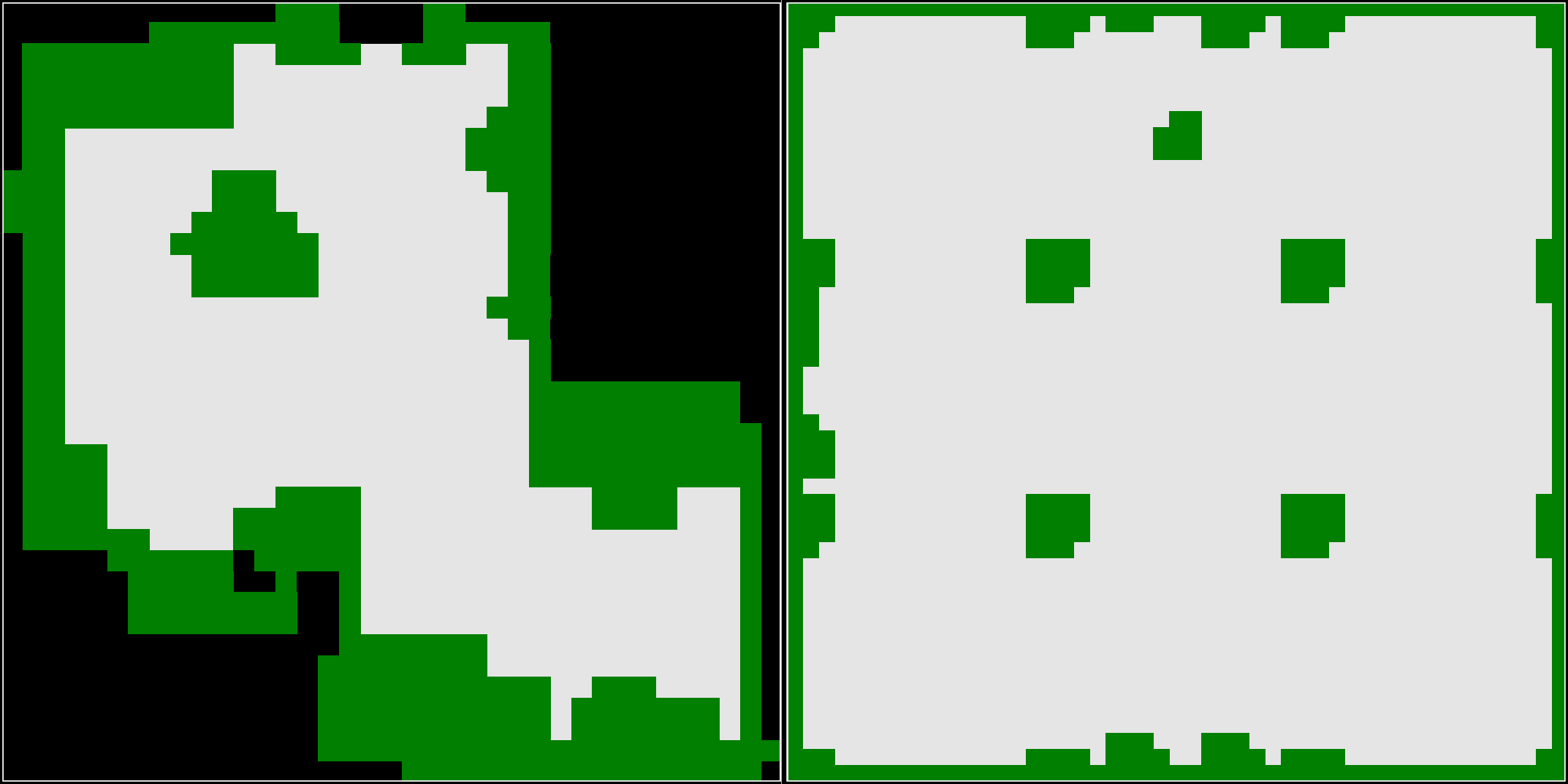}
\caption{Game maps: (left) den201d and (right) arena. }
\label{videogamemaps}
\end{figure}

We conducted a series of experiments (on a 5.0 GHz Intel Core i9-9900K computer with 16GB RAM) across benchmark maps~\cite{stern2019mapf}, with multiple parameter configurations. We ran a test suite with permutations of the three warehouse-like maps (small, medium, and large shown in Figure~\ref{warehouse_maps}), two video game maps (shown in Figure~\ref{videogamemaps}), and parameters: controlled agent set sizes of $[5, 10, 20, 30, 40, 50]$, uncontrolled agent set sizes of $[1, 3, 5, 10, 20, 30, 40]$, conformal confidence level $\delta := \{0.01, 0.05\}$,  conflict horizon $\hat{w} := \{1, 3, 5, 10, 15\}$ and replanning window $H := \{1, 3, 5, 10, 15\}$ for which $H \leq \hat{w}$. 

We compare our approach against three other algorithms. First, IGNORE-ECBS, which uses the standard ECBS solver with no knowledge of dynamic agents to produce a baseline~\cite{barer2014suboptimal}. Second, OBSTACLE-ECBS, sensing the current locations of agents and treating them as static obstacles for the entire planning window. Third, PRED-ECBS, a version of CP-Solver that uses predictions, but not CP intervals.
\begin{figure*}[t]
\centering
\includegraphics[width=0.86\textwidth]{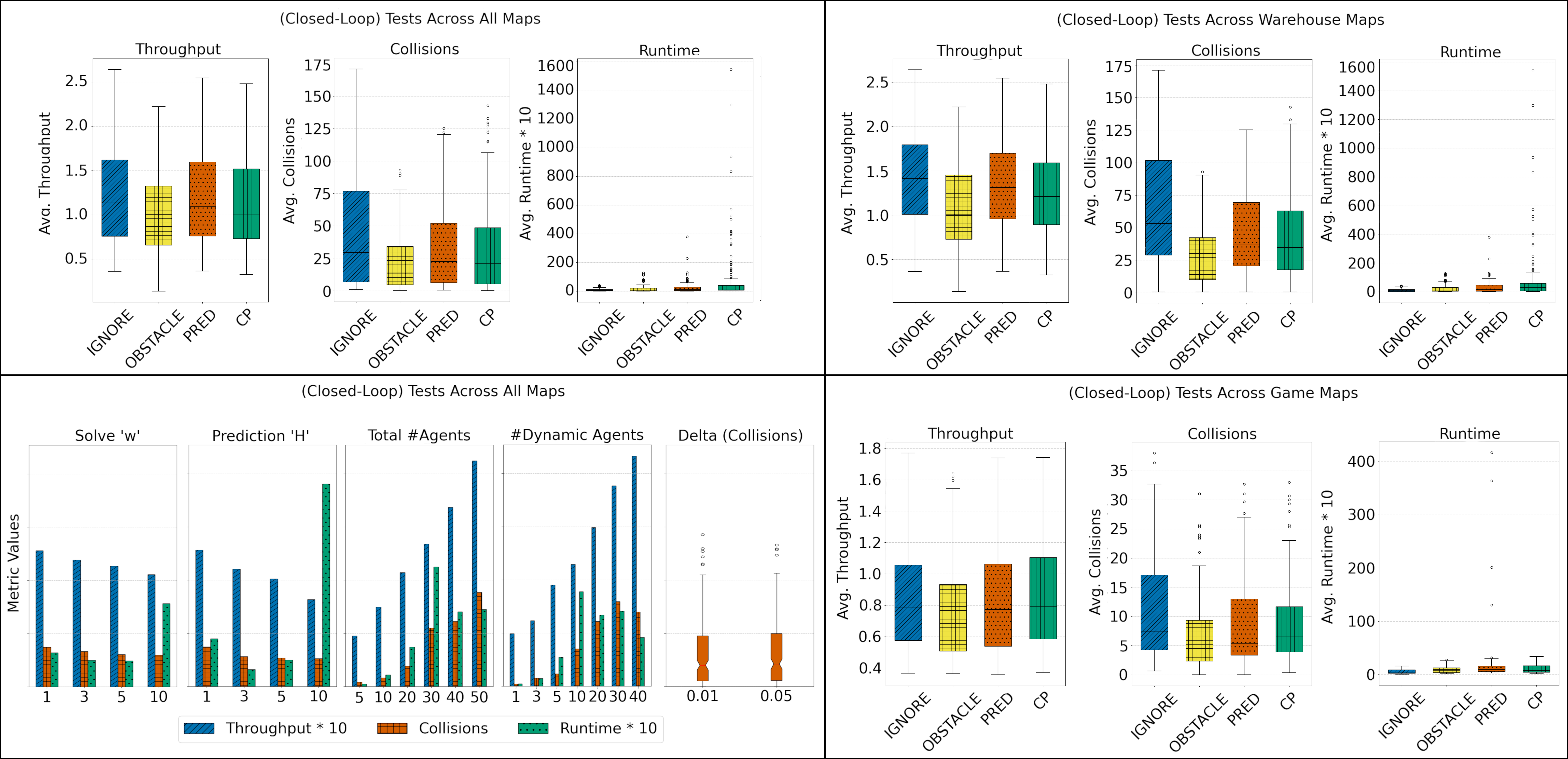}
\caption{Closed-loop CP-Solver averaged across all agent initialization, maps, and configuration parameters not specified: (upper-left) all maps, (upper-right) warehouse maps, (lower-right) video game maps, (lower-left) CP-Solver.}
\label{results:closed_loop}
\end{figure*}
Results were averaged across three random initializations of agent starting positions and goal assignments for each permutation of configuration parameters to account for variability in initial placements. We recorded several performance metrics for each test run: 1) throughput for goal attainment performance, 2) runtime for solver feasibility, and 3) number of collisions with uncontrolled agents for safety. We note that collisions are counted at each timestep; if two agents remain in a collision, they will increase the collision count at every timestep.

We collected $5,000$ trajectories of dynamic agents for each map and dataset type operating with A* paths to random goals. We trained a simple LSTM network for each map type and size, setting the architecture to be $2$ layers of $128$ hidden units. Our model uses a history of length $4$ with various prediction horizons of $(1, 3, 5, 10)$ timesteps to predict the path of dynamic agents. We assign the closest unassigned task spot to the final prediction as the goal vertex and assume the agent stays in this goal location. We set the ECBS suboptimality bound to $1.5$, aligned with the libMultiRobotPlanning library~\cite{honig2016multi}. This library implements popular task and path planning algorithms. We have modified their version of ECBS and increment the focal weight by $1$ after a $100$-second timeout.

\section{Results}
We illustrate the closed-loop results in Figure~\ref{results:closed_loop}. 
\begin{table}[t]
\centering
\begin{tabular}{|l|c|c|c|c|c|}
\hline
\textbf{Map} & \textbf{Small} & \textbf{Normal} & \textbf{Large} & \textbf{Arena} & \textbf{Den201d} \\
\hline
\textbf{Rate} & 0.978 & 0.987 & 0.962 & 0.981 & 0.984\\
\hline
\end{tabular}
\caption{CP coverage of trajectory prediction error across 100 open-loop instances per map, with $\delta = 0.05$.}
\label{tab:map_rates}
\end{table}

\begin{table}[t]
\centering
\begin{tabular}{|l|c|c|c|c|c|}
\hline
\textbf{Uncontrolled Agents} & \textbf{1} & \textbf{3} & \textbf{5} \\
\hline
\textbf{Safety Violations} & 0/100 & 0/100 & 2/100\\
\hline
\end{tabular}
\caption{Number of the $100$ \textbf{open-loop} CP-Solver runs that have one or more collisions (a safety violation) with parameters: \#controlled $:= 10$, $\delta := 0.05$, and $H := 15$.}
\label{tab:open_rates}
\end{table}

\begin{table}[t]
\centering
\begin{tabular}{|l|c|c|c|c|c|}
\hline
\textbf{Uncontrolled Agents} & \textbf{1} & \textbf{3} & \textbf{5} \\
\hline
\textbf{Safety Violations} & 1/100 & 3/100 & 4/100\\
\hline
\end{tabular}
\caption{Number of the $100$ \textbf{closed-loop} CP-Solver runs that have one or more collisions (a safety violation) with parameters: \#controlled $:= 10$, $\delta := 0.05$, $H := 10$, and $\hat{T} := 100$.}
\label{tab:closed_rates}
\end{table}

\textbf{Probabilistic Collision Avoidance}: Corollary~\ref{cpcorollary} ensures safe planning around dynamic agents at a user-defined confidence level. With $\delta=0.05$, Table~\ref{tab:map_rates} displays the results of running our open-loop formulation across all maps and parameters,  confirming that our CP intervals capture $95\%$ of prediction errors in the prediction window. In Tables~\ref{tab:open_rates} and ~\ref{tab:closed_rates}, we show that the number of CP-Solver runs, for a selection of parameters and agents, that have a safety violation (one or more collisions) is less than the expected $\delta$ bound aligning with Corollary \ref{cpcorollary}. Assumption~\ref{interference_assumption} acknowledges CP's vulnerability to distribution shifts, although the theoretical guarantees hold under small shifts~\cite{cauchois2023robust,strawn2023conformal} and robust CP extensions exist for larger shifts~\cite{angelopoulos2024theoretical,cauchois2023robust}.  

We note that OBSTACLE-ECBS produces fewer collisions on average; however, this is due to the increased number of obstacles that deadlock areas of the map, forcing agents to wait or take alternative paths. Differences in map and agent configurations can significantly impact the performance of OBSTACLE-ECBS. CP-Solver offers guarantees for reliable performance and safety, using predictions to enable more agents to keep moving in the direction of their goal, which may bring them closer to uncontrolled agents than OBSTACLE-ECBS. 

\textbf{Throughput}: CP-Solver achieves higher throughput than OBSTACLE, closely trailing PRED (lacks guarantees) and IGNORE (disregards dynamic agents entirely). 

\textbf{Runtime}: CP-Solver runtime remains competitive, though closed-loop runs occasionally exhibit higher outliers. Some initializations lead to increased runtimes, as CP-Solver retains more agents in the solver while IGNORE and OBSTACLE exclude more agents. Future work could explore ECBS node expansion and runtime optimizations.

\textbf{Horizon Parameters}: As $\hat{w}$ and $H$ increase, collisions decrease slightly while throughput and runtime increase. CP-Solver throughput decreases and runtime increases at large $H$ values, consistent with findings that extended horizons in RHCR do not always improve overall performance~\cite{li2021lifelong}. Improving the predictor has a more significant impact on performance than greatly increasing prediction length. 

\section{Conclusion}
 Presenting the MAPF DUA problem and CP-Solver broadens the scope of CBS-based solvers and provides statistical safety guarantees for MAPF applications in environments with uncontrolled dynamic agents. We have contributed CP-Solver, which reduces collisions compared to standard approaches and remains competitive in terms of throughput, number of collisions, and runtime across experiments in warehouse and video-game environments. CP-Solver provides an approximate solution for one-shot and lifelong MAPF among uncontrollable agents, enabling path planning with safety guarantees by leveraging uncertainty quantification with predictions and adapting Enhanced Conflict-Based Search. Future work could focus on robustness to distribution shifts or enhancing prediction accuracy. 

\bibliography{aaai2026}

\end{document}